\documentclass[12pt,preprint]{aastex}

\begin{document}

\title{Constraints on the TeV source population and its contribution to the galactic
diffuse TeV emission}

\author{S. Casanova \footnote{current address: ICRA, Universita' degli Studi La Sapienza di Roma, 00185 Roma, Italy and Max-Planck-Institut für Kernphysik
Saupfercheckweg 1 69117 Heidelberg, Germany} and B. L. Dingus}
\affil{Los Alamos National Laboratory, Los Alamos, NM 87545, US}
\begin{abstract}

The detection by the HESS atmospheric Cerenkov telescope of fourteen new
sources from the Galactic plane makes it possible
to estimate the contribution of unresolved sources like those detected by HESS to
the diffuse Galactic emission measured by the Milagro Collaboration. The number-intensity relation and the
luminosity function for the HESS source population are investigated.
By evaluating the contribution of such a source population to the diffuse emission we conclude
that a significant fraction of
the TeV energy emission measured by the Milagro experiment could be due to unresolved sources like
HESS sources. Predictions
concerning the number of sources which Veritas, Milagro, and HAWC should detect are also given.

\end{abstract}

\section{Introduction}

The Galactic diffuse $\gamma$-ray emission is believed to be mostly produced in interactions
of cosmic rays with the matter and the radiation
fields in the Galaxy, the main production mechanisms being electron non-thermal Bremsstrahlung,
Inverse Compton scatterings
off the radiation fields and pion decay processes in inelastic collisions of nuclei and matter.
Although the standard production mechanisms of $\gamma$-rays \citep{Bertsch:1993} explain
generally well the spatial and
energy distribution of the emission below 1 GeV,
the model does not match EGRET observations of the $\gamma$-ray sky above 1 GeV \citep{Hunter:1997we}.
Many possible explanations have been proposed to account for the GeV excess \citep{Aharonian:2000iz,
Strong:1999hr,Strong:2004,Kamae:2006, deBoer:2005tm, Bergstrom:2006tk,Strong:2006,Stecker}. In any case a major difficulty when
studying the diffuse emission is to disentangle the truly diffuse emission from that produced by unresolved sources.

The TeV diffuse $\gamma$-ray emission from the Galactic Plane \citep{Atkins:2005} and from the Cygnus region \citep{Abdo:2006}
has been measured by Milagro, a water Cerenkov telescope surveying the northern sky at
TeV energies \citep{Atkins:2000,Atkins:2003}. The diffuse emission measured by Milagro is $(7.3  \pm 1.5  \pm 2.3)  \times {10}^{-11} \, photons
 \, {\mathrm{{cm}^{-2}}} \,{\mathrm{{sr}^{-1}}} \, {\mathrm {s^{-1}}} $ for $E>3.5$ TeV
in the region ${40}^{o}<l<{100}^{o}$ \citep{Atkins:2005}. A measurement of the TeV $\gamma$-ray diffuse flux is important, as it
allows us to calculate what fraction of the emission is produced through
inverse Compton and what fraction is produced through pion-decay mechanism. Milagro result, which is consistent
with EGRET data above 1 GeV, seems to exclude an additional hard spectrum component
continuing to above 10 TeV to explain the 60 per cent excess of EGRET flux compared to $\pi_{0}$-decay
production mechanism due to the local cosmic ray flux \citep{Hunter:1997we}. Recently \citet{Prodanovic}
have argued that no strong signal of pion decay is seen in the $\gamma$-ray spectrum. The pion decay
mechanism would not be able to explain the excess above 1 GeV and the Milagro measurements of
the diffuse flux at higher energies reveals a new excess at TeV energies. \citet{Prodanovic}
investigated several possibilities for what might cause the TeV excess, among them possible dark matter decay,
contribution from unresolved EGRET sources or sources that are only bright in the TeV range.
An extrapolation of the EGRET sources within the range in
Galactic longitude of the Milagro observation overestimates the diffuse TeV flux measured.
However, there could be a population of sources undetectable by EGRET, but contributing to both the GeV and TeV excess
diffuse emission. Recent observations by the TeV observatory, HESS, in fact point to a new class of hard spectrum
TeV sources located close to the Galactic Plane, within $\pm 1$ degrees of latitude, whose average slope is about
$E^{-2.3}$ \citep{Aharonian:2005jn,Aharonian:2005kn}. The HESS detection of high energy $\gamma$ rays from fourteen
new sources has improved significantly the knowledge of both the spatial distribution and the spectra and fluxes of
VHE $\gamma$-ray galactic sources. These HESS results make it possible to estimate
with unprecedented precision the contribution of unresolved sources to the Galactic diffuse emission recently extended
to TeV energies by Milagro. In fact, the Milagro sensitivity to point sources is about ${10}^{-11}  \,
photons \, {\mathrm{{cm}^{-2}}}  \, {\mathrm {s^{-1}}} \, {\mathrm for} \quad E > 1 TeV$  and it
is of the order of the brightest integral source flux detected by HESS. Sources like the ones detected by HESS are then
unresolved by Milagro and contribute to the diffuse emission Milagro measures.

\begin{figure}[ht]
\begin{center}
\includegraphics[angle=0,width=8cm]{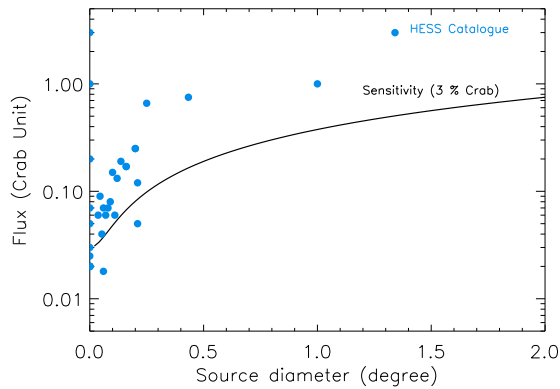}
\caption{The HESS source fluxes and the HESS survey sensitivity versus source extensions are plotted \cite{Aharonian:2005kn}.
A clustering of the sources about the HESS sensitivity limit is evident.}\label{fig1}
\end{center}
\end{figure}

Here, based on HESS results, knowing the sensitivity and the field of view of an experiment, we estimate the
number of expected sources and their expected VHE $\gamma$-ray flux. We then evaluate the contribution of unresolved
sources to the diffuse $\gamma$-ray emission for Milagro. The estimate of the diffuse emission
arising from unresolved sources will be a lower limit. In fact, HESS has mostly detected extended
sources. For sources larger than the angular resolution of the experiment the source size determines the
sensitivity of the instrument. The sensitivity decreases if the extension of the source increases.
In Fig.~\ref{fig1} the HESS source fluxes and the sensitivity as a function of
the source extensions are plotted. A clustering of sources about HESS survey sensitivity of
3 percent of the Crab flux reported in \citep{Aharonian:2005jn} is evident,
meaning that there are probably many more to be detected with better sensitivity.

\section{Galactic spatial distributions of PSRs and SNRs and their VHE counterparts}

Hereafter the following notations will be used.
In terms of the heliocentric distance $D$, the longitude $l$ and the latitude $b$
the galactocentric distance, $r$, is defined as
\begin{equation}
r = \sqrt{z^2 + (r_0^2 + D^2 - 2 \, r_0 \, D \, cos(l))}
\end{equation}
where $r_0=8.5$ ${\mathrm{kpc}}$ is the distance from the Earth to the Galactic center.
The height $z$ over the Galactic Plane is
\begin{equation}
z = D \, sin(b)
\end{equation}

In the inner Galaxy region observed by HESS the existence of 91 SNRs and 389 pulsars was
already known at lower energies \citep{Green:2004gr,Manchester}. Many of these
SNRs and pulsars within the inner Galaxy might emit VHE $\gamma$ rays but only a few $\gamma$ ray sources were
previously known.
Although the HESS Collaboration indicates a clear positional coincidence of its sources with a known SNR or pulsar
only in a limited
number of cases, the distribution of Galactic latitude of
the seventeen VHE $\gamma$ ray sources detected by HESS agrees quite well with the distributions of all SNRs catalogued by
\cite{Green:2004gr} and of all pulsars catalogued by \cite{Manchester}. Assuming, for example, that SNRs are a single class of counterparts
with isotropic luminosity $1.95  \times {10}^{34}  \,  {\mathrm{erg}}  \,  {\mathrm{s^{-1}}}$
to the new HESS sources and taking for them a simple radiative model, \cite{Aharonian:2005kn} found that the location of these sources
favours a scale height of less than 100 pc, consistent with the hypothesis that these sources are either SNRs or pulsars in
a massive star forming region. Therefore although for only a few of HESS sources a firm identification with counterparts
at other wavelengths exists, there are some suggestions that many of the HESS
sources might coincide with supernova remnants (SNRs) or pulsar wind nebulae (PWNe).
In fact two of the HESS sources have SNRs as counterparts, and five of
these most recently discovered HESS sources are associated with pulsar wind nebulae \citep{Funk}.
SNRs are an established source class in VHE $\gamma$ ray astronomy. Possible correlations between
SNRs and unidentified EGRET and HESS sources have been proposed since the release of the first EGRET
catalogue \citep{Sturmer,Esposito}, and later for the third EGRET catalogue by \citep{Romero,Combi}.
PWNe formed from young pulsars with age less than a million years are considered as
potential gamma-ray emitters \citep{Manchester:2004}. Though a young age is not a sufficient
condition for a pulsar to generate a PWN. The spin-down energy loss is the key parameter
to determine whether a young energetic pulsar forms a PWN \citep{Gotthelf:2003}.  The ratio
between $\gamma$-ray loud versus $\gamma$-ray quiet pulsars is uncertain. \citet{Gotthelf:2003}
suggests that all pulsars with $dE/dt > dE/dt_c = 3.4 \times {10}^{36} erg/s$
are X-ray bright, manifest a distinct pulsar wind nebula (PWN), and are associated with a supernova
event. By studying the Chandra data on the 28 most energetic pulsars
of the Parkes Multibeam Pulsar Survey \citep{Manchester:2004} \citet{Gotthelf:2004} found that
15 pulsars with $\dot E > 3.4 \times {10}^{36} ergs/s$ are X-ray bright,
show a resolved PWN, and are associated with evidence of a supernova event. This suggests that about
2.5 per cent of the radio loud pulsar have a PWN and might emit $\gamma$-rays.

\begin{figure}[ht]
\begin{center}
\includegraphics[angle=0,width=8cm]{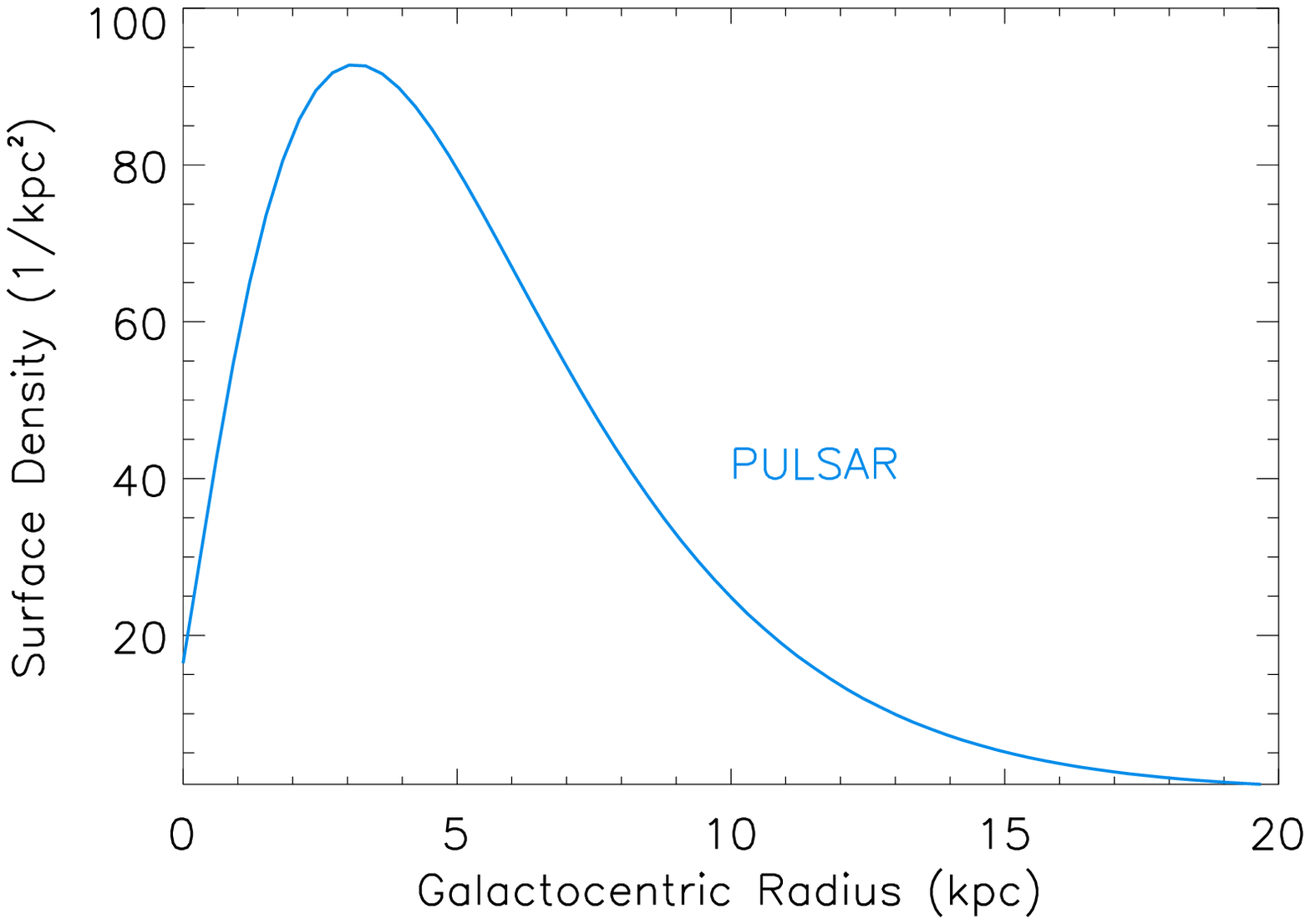}
\includegraphics[angle=0,width=8cm]{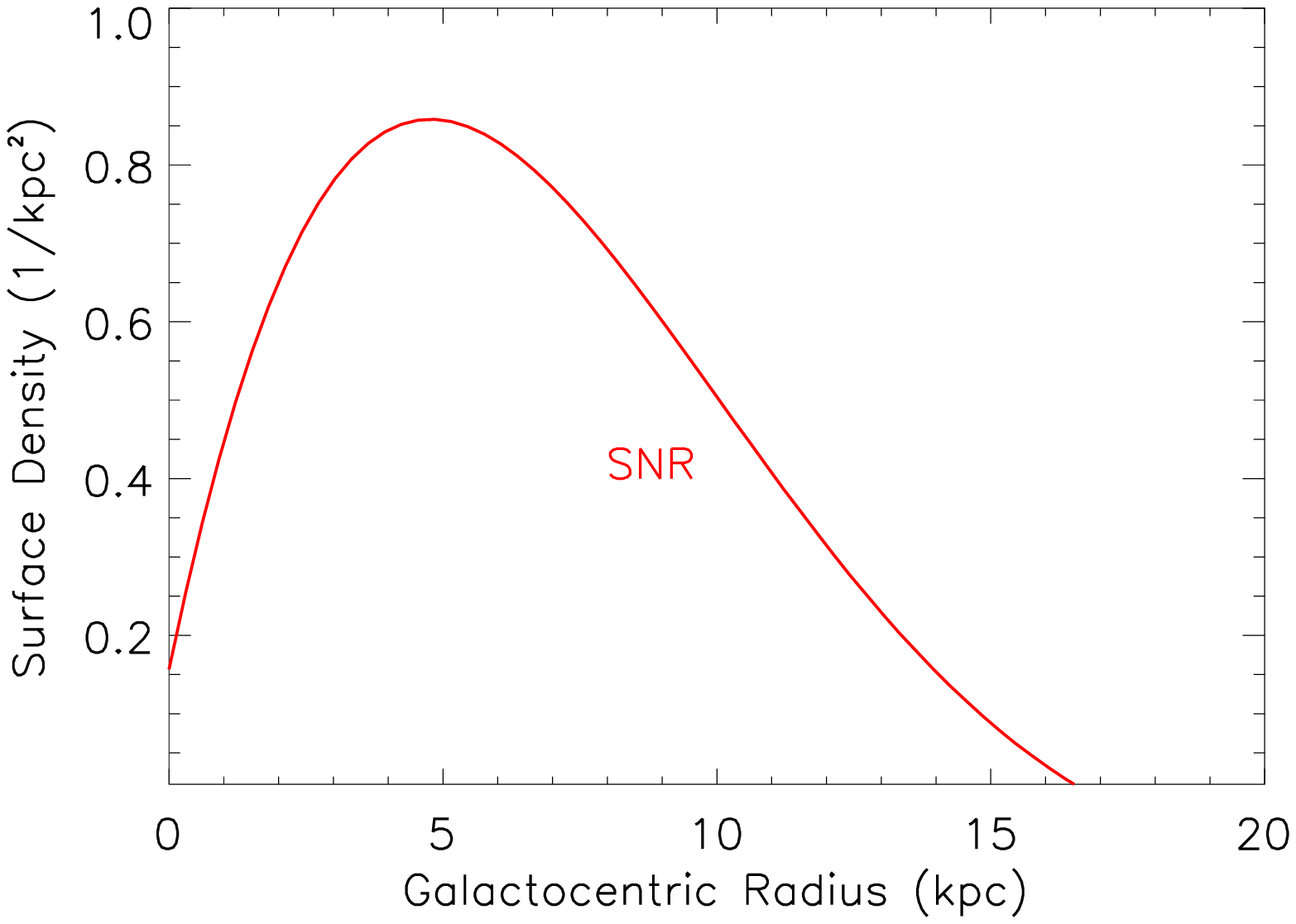}
\caption{{ Pulsar and SNR surface distributions versus galactocentric radius $r$ from
\citet{Lorimer:2006} and from \citet{Green:2004gr} and \citet{Case:1998qg}, respectively.} \label{fig2}}
\end{center}
\end{figure}

Supernova remnants and pulsars are the radio counterparts of two of the high energy
gamma ray candidates, SNRs and PWNe, and their spatial distribution is known from many
observations at radio wavelengths. The pulsar surface density $\sigma_{PSR}(r)$,
plotted in Fig.~\ref{fig2}, is fitted by
the following shifted Gamma function \citep{Yusifov:2004fr,Lorimer:2004,Lorimer:2006}
\begin{equation}
\sigma_{PSR}(r,z) = a \,{(\frac{r}{r_0})}^b \, e^{[-c \, (\frac{r-r_0}{r_0})]}
\end{equation}
where $a=41$ ${\mathrm {kpc^{-2}}}$ and $b=1.9$ and $c=5.0$. For the $z$ distribution we use the
exponential function
\begin{equation}
n_{PSR}(z) = d\,  e^{-\frac{|z|}{e}}
\end{equation}
where $d=0.75$ and $e=0.18{\mathrm{kpc}}$ \citep{Lorimer:2006}.
The SNR surface density, plotted in Fig.~\ref{fig2}, is
\citep{Green:2004gr,Case:1998qg}
\begin{eqnarray}
\sigma_{SNR}(r) = \left\{ \begin{array}{c@{\hspace{12mm}}l}
{\sigma_0}_{SNR}  \,sin(\frac{\pi \,r}{r_2}+\theta_0) \,  e^{-\beta r}  &  \quad {\mathrm for}
\quad {\mathrm r} < 16.8 \\[2mm] 0 &  \quad {\mathrm for} \quad {\mathrm r} > 16.8
\end{array} \right.
\label{sigmasnr}
\end{eqnarray}
with ${\sigma_0}_{SNR} = 1.96\pm 1.38$ ${\mathrm {kpc^{-2}}}$, $r_2=17.2\pm 1.9$ ${\mathrm {kpc}}$,
$\theta_0=0.08\pm0.33$ and $\beta=0.13\pm0.08$. For the $z$ distribution we use the exponential function
\begin{equation}
n_{SNR}(z) = d\,  e^{-\frac{|z|}{e}}
\end{equation}
where $d=0.58$ and $e=0.083{\mathrm{kpc}}$ \citep{Xu}.
Assuming the SNR and PSR distributions plotted in Fig.\ref{fig2}, the Galaxy should contain
217 SNRs and 13176 PSRs.
\begin{figure}[ht]
\begin{center}
\includegraphics[angle=0,width=8cm]{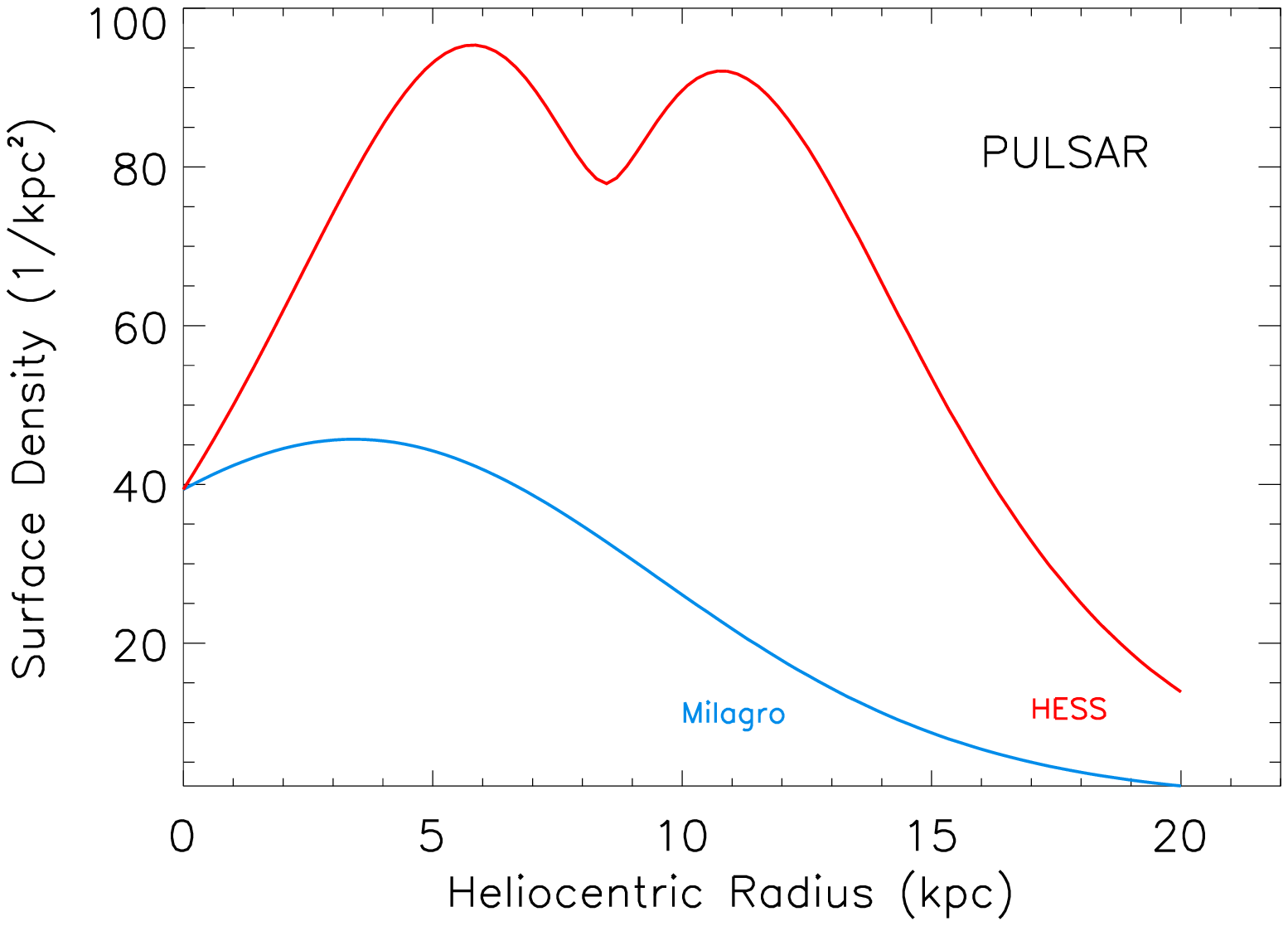}
\includegraphics[angle=0,width=8cm]{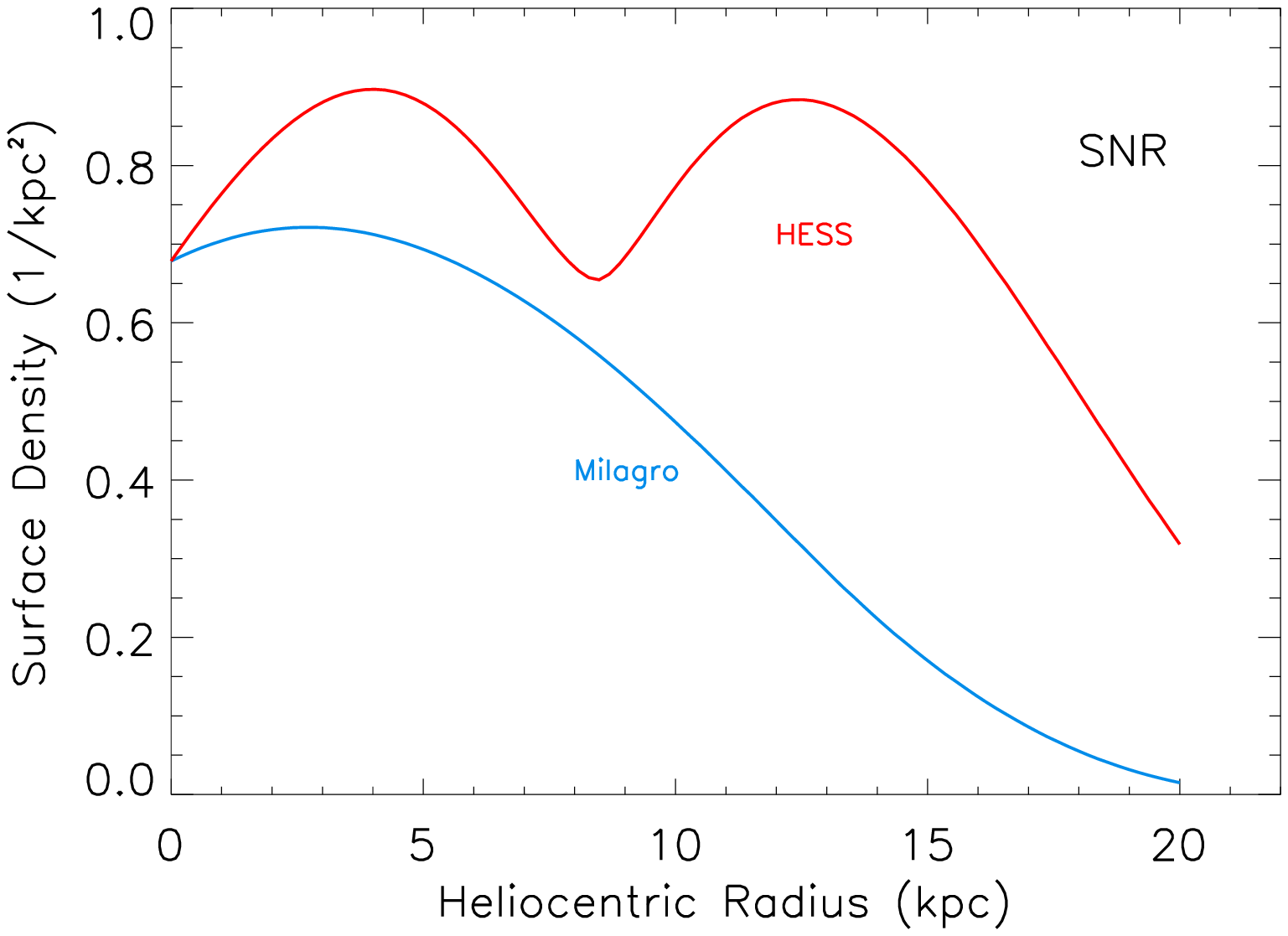}
\caption{Pulsar and SNR surface distributions versus the heliocentric distance D for the regions
of the Galactic plane surveyed
by Milagro (~${40}^{o}<l<{100}^{o}$~) and by HESS (~$-{30}^{o}<l<{30}^{o}$~).\label{fig3}}
\end{center}
\end{figure}

\section{Number-intensity relation for HESS sources}
In order to perform a study of the collective properties of HESS source population
the number-intensity relation, $log N (>S)-log S$, is
here used. The number-intensity relation has the advantage of using the flux data
without any assumption on the distance and
luminosity. In fact for many of HESS sources the location and luminosity are unknown.
The number-intensity relation also gives information on the geometry of the volume
in which the sources are contained if a uniform source distribution is assumed. Deviation
from a simple power law $N \propto S^{-\alpha}$ with slope -1 (for the case of a thin disk) means that
the sources or their luminosity function are not spatially uniformly distributed.

\begin{figure}[ht]
\begin{center}
\includegraphics[angle=0,width=8cm]{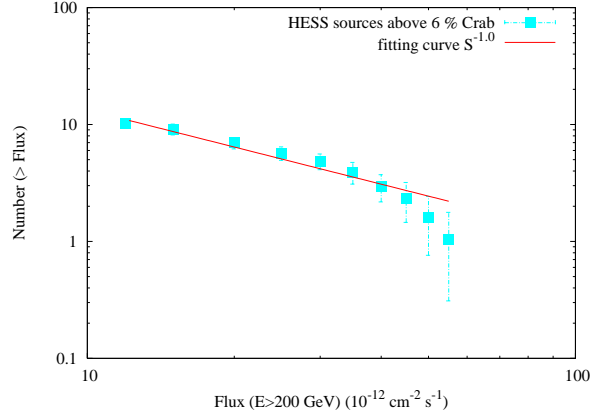}
\caption{The number-intensity relation, $log N (>S)-log S$, for HESS source population. In order to
have a complete sample, only sources detected above 6 per
cent of the Crab flux are included\label{fig4}. The population of sources is distributed in a thin disk.}
\end{center}
\end{figure}

The major difficulties in the study of the collective properties of the
HESS source population consist of the limited number of sources detected and the
relatively small range of flux covered by the survey. Also, the HESS survey of the Galaxy
was not performed with uniform sensitivity.
In fact, whereas the sensitivity of the survey of the Galactic Plane in Galactic latitude
is rather flat in the region between -1.5 and 1.5 degrees, its effective exposure and therefore
its sensitivity is not uniform in longitude. Longer observation
times were dedicated by HESS to locations in the Galactic plane close to where three sources,
HESS J1747-218 and HESS J1745-290 (in the Galactic Center), and HESS J1713-397, were already known.
The average sensitivity of the survey as a function of the longitude and the latitude are
shown in Fig.~2 and Fig.~3 of \cite{Aharonian:2005kn}, respectively. In some locations of the Galaxy
the survey was done at peak sensitivity of 2 per cent of the Crab flux. From Fig.~3 of
\cite{Aharonian:2005kn} one can deduce that in order for our sample to be complete, only sources
detected with more than 6 per cent the Crab flux within $-{2}^{o}<l<{2}^{o}$ can be included. Table
1 lists the eleven sources included to plot the number-intensity relation.
\begin{table}[ht]
\begin{center}
\begin{tabular}{|c|c|}
\hline
SOURCE NAME    &  FLUX ABOVE 200 GeV     \\[2mm] \hline
$HESS J1614-518$     &  $57.8\pm 7.7$    \\[2mm]
\hline
$HESS J1616-508$     &  $43.3 \pm 2.0$    \\[2mm]
\hline
$HESS J1632-478$     &  $28.7\pm 5.3$     \\[2mm]
\hline
$HESS J1634-472$     &  $13.4\pm 2.6$     \\[2mm]
\hline
$HESS J1640-465$     &  $20.9\pm 2.2$     \\[2mm]
\hline
$HESS J1702-420$     &  $15.9\pm 1.8$     \\[2mm]
\hline
$HESS J1804-216$     &  $53.2\pm2.0$     \\[2mm]
\hline
$HESS J1813-178$     &  $14.2\pm1.1$     \\[2mm]
\hline
$HESS J1825-137$     &  $39.4\pm 2.2$    \\[2mm]
\hline
$HESS J1834-087$     &  $18.7\pm 2.0$    \\[2mm]
\hline
$HESS J1837-069$     &  $30.4\pm 1.6$    \\[2mm]
\hline
\end{tabular}
\caption{HESS sources included in the logN-logS with fluxes above 6 per cent of the Crab flux.
All fluxes for $E>200$ GeV are in units of in ${10}^{-12}\, photons
 \, {\mathrm{{cm}^{-2}}} \, {\mathrm {sr}^{-1}} \, {\mathrm {s^{-1}}} $ \cite{Aharonian:2005kn}.
\label{Table1}}
\end{center}
\end{table}
HESS source fluxes have small statistical errors, whereas their systematical uncertainties are
about 30 per cent of the absolute value of the flux. In order to account for the
large systematic uncertainties in the fluxes a Monte-Carlo based procedure is adopted to
generate the errors for the $log N(>S)-log S$ diagram, plotted in Fig.~\ref{fig4}. A random number
spanning an interval equal to the systematic and statistical errors in quadrature around the
flux value is generated. By repeating the procedure many times standard deviations can be
evaluated \citep{Bignami}. These standard deviations are the error bars in the $logN-logS$
diagram plotted in Fig.~\ref{fig4}. The power law fitting curve has a slope
$-1.0\pm 0.1$ with reduced $\chi^2=0.5$ for the HESS sample for integral fluxes bigger than
$S_{1} = 12 \times {10}^{-12} \, {{cm^2}} \, {s^{-1}}$, which corresponds to 6 per cent of the Crab flux.
The number-intensity relation is
\begin{equation}
N(>S) = (152 \pm 41)\, {(\frac{S}{S_0})}^{(-1.0\pm 0.1)}
\end{equation}
where $S_0= {10}^{-12} \, {{cm^2}} \, {{s^{-1}}}$. The reduced $\chi^2$ for a fit with slope -1.5 would be 3.3.
By assuming a tridimensional
isotropic distribution of sources the slope of the number-intensity relation should
be -1.5. The fact that the slope is about -1 means that the source decrease in the
flux with distance is counterbalanced by the density of sources depending on the inverse square of
the distance. The HESS-like sources are distributed in a thin disk, in a volume
that is larger than the visibility limit.
\subsection{Predictions}
The $\gamma$-ray flux due to the HESS source population above 6 per cent
of the Crab flux in the region $-{30}^{o}<l<{30}^{o}$ and $-{2}^{o}<l<{2}^{o}$ is
\begin{equation}
F(E> 200 GeV, \,  -30<l<30, \, -2<b<2) = \int_{S_1}^{S_2}  N(>S) \, dS =   2.5
\times {10}^{-10} \, photons\,  {\mathrm{{cm}^{-2}}} \, {\mathrm {s^{-1}}} \,,
\label{eqnfluxtotal}
\end{equation}
where $S_2= 57.8 \times {10}^{-12} \, {{cm^2}} \, {s^{-1}}$
is the maximum flux detected by HESS from a source.

From the PSR and SNR distributions plotted in Fig.(\ref{fig2})
there are 84 SNRs and 5324 PRSs in the region $-{30}^{o}<l<{30}^{o}$ and $-2^o<b<2^o$ , whereas in
the Milagro region (${40}^{o}<l<{100}^{o}$ and $-5^o<b<5^o$) there are 36 SNRs and 1358 PSRs.
Assuming that supernova remnants and pulsars are the radio counterparts of high energy
gamma ray sources, SNRs and PWNe, the Milagro region has 26 percent of the
sources which are in the HESS region. Also, the Milagro Galactic diffuse emission is measured
for a threshold energy of about 3.5 TeV, whereas the flux from unresolved sources
calculated in Eq.(\ref{eqnfluxtotal}) refers to the HESS threshold energy of 200 GeV.
In order to estimate the contribution of unresolved sources to Milagro diffuse emission
the flux in Eq.(\ref{eqnfluxtotal}) has to be corrected
for the different threshold energy. All HESS sources are fitted by a power law spectrum
\begin{equation}
\Phi(E,\Gamma) = \Phi_0 \, {( \frac{E}{1 TeV} )}^{-{\Gamma}} \,,
\end{equation}
and the average spectral index is $\Gamma=2.32$. By correcting the flux in Eq.(\ref{eqnfluxtotal}) for
the Milagro threshold energy the flux which HESS-like sources contribute is
\begin{eqnarray*}
F(E>3.5 \, TeV, \, 40<l<100,\, -5<b<5 ) &=&  0.26 \, F(E>200 \, GeV, \, -30<l<30) \\[2mm]
& \times &{(\frac{E_{Milagro}}{E_{HESS}})}^{(-\Gamma+1)} \\[2mm]
&& =  8.3 \times {10}^{-12} \, photons  \,{\mathrm{{cm}^{-2}}} {\mathrm {sr}^{-1}} \, {\mathrm {s^{-1}}}\,.
\end{eqnarray*}
\begin{equation}
\label{fluxtotaltris}
\end{equation}
The contribution of HESS source population amounts to at least
10 per cent of the diffuse flux which Milagro measures above 3.5 TeV. This is a lower limit for
the contribution of unresolved sources to Milagro diffuse emission, as only sources above
6 percent of the Crab flux were taken into account to estimate it.

The Galactic diffuse emission measured by Milagro can be used to constrain the minimum flux $S_{min}$
below which the logN-logS plot becomes flat in order not to overproduce the Milagro flux
\begin{equation}
\int_{S_{min}}^{S_2} \, dS \,  \frac{dN}{dS}  < 7.3  \times {10}^{-11} \, photons  \,  {\mathrm{{cm}^{-2}}}\, {\mathrm{{sr}^{-1}}} \, {\mathrm {s^{-1}}} \,.
\label{eq:constraint}
\end{equation}
$S_{min}$ cannot be less than $1 \times {10}^{-16} \, photons \, {\mathrm{{sr}^{-1}}} {\mathrm{{cm}^{-2}}}  \,
{\mathrm {s^{-1}}}$ in order not to violate the constraint in
Eq.(\ref{eq:constraint}).

From the number-intensity relation for HESS source population it is possible to deduce
the number of sources which HESS, VERITAS, Milagro and HAWC will detect. The number of SNRs and PWNs
expected for HESS if its entire
field of view is scanned with a uniform sensitivity of 2 per cent of the Crab flux above 200 GeV is about
$43\pm 10$. If VERITAS will survey the Northern sky in the region between ${30}^{o}<l<{220}^{o}$
and ${-10}^{o}<b<{10}^{o}$ reaching the level of 1 per cent of the Crab flux above 100 GeV, it should detect approximately $35\pm10$ sources.
Milagro has previously observed the Northern sky with a sensitivity equal to about 65
per cent the Crab flux at 3.5 TeV. At this level of sensitivity one should expect no detection of sources
for Milagro and indeed no detection of sources was claimed in \citet{Atkins:2005} with the data
accumulated after the first three years of operation. Now after six years of data have been accumulated
Milagro has reached a sensitivity equal to about 20 per cent of the Crab flux at 20 TeV median energy and
should be able to detect $3\pm1$ sources. The Milagro Collaboration has recently published the survey of the Northern
sky in the region between  ${30}^{o}<l<{220}^{o}$ and  ${-10}^{o}<b<{10}^{o}$ at a threshold of 20 TeV and
a threshold sensitivity of about 20 percent of the Crab detecting 4 sources \cite{Abdo:2007}. The number of sources detected is in agreement
with our predictions. After only one year of operation the proposed experiment HAWC
will have surveyed the region of the sky with longitude ${5}^{o}<l<{110}^{o}$ and ${130}^{o}<l<{250}^{o}$
and latitude ${-10}^{o}<b<{10}^{o}$ at 30mCrab sensitivity above 1 TeV and should have
detected $19\pm5$ HESS-like SNRs and PWNe. The predictions given here for Milagro and HAWC are valid unless
the source spectra steepen or cut-off below the Milagro threshold. However, for the sources detected by HESS the spectrum
is predominantly well characterized by a single power law.

\section{Alternative method to estimate the contribution of unresolved sources to the diffuse emission}

Thanks to the logN-logS relation we obtained a lower limit for the contribution of unresolved
HESS-like sources to the diffuse emission measured by Milagro. Here we will assume that the
density of VHE $\gamma$-ray source candidates, SNRs and PWNe, follows the volume density of SNRs or
of pulsars in the Galactic Plane as observed at radio wavelengths and their luminosity function is a
power law, and we will then estimate their contribution to the Milagro diffuse emission.

Since the data on HESS sources are too sparse to constrain their luminosity function, we will leave
it as a parametrised input. The luminosity function $\Phi(L)$ will be a power law with different
indices $\alpha$ varying between -1 and -2
\begin{equation}
\Phi(L)= \frac{dN}{dL_{\gamma}}= c\,{(\frac{L_{\gamma}}{{L_{\gamma}}}_0)}^{\alpha}\,.
\label{eqn:luminosity}
\end{equation}
The assumed luminosity function will then be compared with the HESS source counts to fix the
normalisation $c$. In Eq.(\ref{eqn:luminosity}) ${L_{\gamma}}_0= 1 \times {10}^{34} erg/s$.
The range in luminosities for the HESS sources, for which
the distance and thus the luminosity
is known, varies between ${10}^{31} erg/s$ and
${10}^{36} erg/s$. In fact, most sources of $\gamma$-ray in the Galaxy are located close
to the plane of the Galaxy, within a region which extends from $D_{min}=0.3$ kpc up to
$D_{max}=30$ kpc \citep{Swordy}. The range in luminosity for the HESS sample can then be found from
the HESS sensitivity (we assume 6 percent of the Crab flux) and the maximum flux detected by HESS,
which are respectively
\begin{eqnarray*}
{L_{\gamma}}_{min} &=& \frac{\Gamma-1}{\Gamma-2} \, {E_{th}} \,  4 \, \pi \, {D_{min}}^2 \, f_{min}
= 3 \times {10}^{31} erg/s \\[2mm]
{L_{\gamma}}_{max} &=& \frac{\Gamma-1}{\Gamma-2} \, {E_{th}} \,  4 \, \pi \, {D_{max}}^2 \, f_{max}
= 1 \times {10}^{36} erg/s
\end{eqnarray*}
\begin{equation}
\label{lumfun}
\end{equation}
where $E_{th}$ is the detector threshold energy and $\Gamma$ is the spectral index if the
$\gamma$-ray emission is a power law.
By requiring that the following integral over HESS field of view and over the luminosity
range given in Eq.(\ref{lumfun}) gives the number of sources HESS detects above 6 percent of the Crab
flux we will obtain the normalisation factor $c$
\begin{eqnarray*}
N({L_{\gamma}}_{min}, {L_{\gamma}}_{max},V) &=&
\int_{V_{HESS}} dV  \, \int_{{L_{\gamma}}_{min}}^{{L_{\gamma}}_{max}} dL \,
\frac{dN_{sources}}{dVdL} = 11 \\[2mm]
&& \int_{V_{HESS}} dV  \, \int_{{L_{\gamma}}_{min}}^{{L_{\gamma}}_{max}} dL \, \rho_{sources}(r,z) \,
c\,{(\frac{L_{\gamma}}{{L_{\gamma}}}_0)}^{\alpha} \\[2mm]
&&\int_{0.3}^{30} dD \, D^2 \, \int_{-30}^{30} d l \, \int_{-2}^{2} d b \, sin(b) \, \rho_{sources}(l,b,D)
\int_{{L_{\gamma}}_{min}}^{{L_{\gamma}}_{max}}
\, dL \,  \, c\,{(\frac{L_{\gamma}}{{L_{\gamma}}}_0)}^{\alpha} \,.
\end{eqnarray*}
\begin{equation}
\label{integral}
\end{equation}
In Eq.(\ref{integral}) $\rho_{sources}$ is the density of SNRs and PWNe.

\subsection{Predictions}

The integral flux above $E_0$ from resolved and unresolved sources
from a region in the Galaxy which extends from $D_{min}$ to $D_{max}$ in
heliocentric distance, from $b_{min}$ to $b_{max}$ in latitude and from $l_{min}$ to $l_{max}$
in longitude is
\begin{eqnarray*}
F(E>E_0, \,  l_{min}<l<l_{max}, \, b_{min}<b<b_{max}) &=& \int_{0}^{{L_{\gamma}}_{max}} \, dL_{\gamma} \,
\Phi(L_{\gamma})\, F(L_{\gamma},D) \, dN_{sources}(r,z) \\[2mm]
&=&   \int_{0}^{{L_{\gamma}}_{max}} dL_{\gamma} \Phi(L_{\gamma}) F(L_{\gamma},D) \int dV
\sigma_{sources}(r) n_{sources}(z) \\[2mm]
&=&   \int_{0}^{{L_{\gamma}}_{max}} \,  dL_{\gamma} \, \Phi(L_{\gamma})\, F(L_{\gamma},D) \int_{D_{min}}^{D_{max}}
dD \, D^2  \\[2mm]
&&  \int_{l=l_{min}}^{l=l_{max}} d l \, \int_{b=b_{min}}^{b=b_{max}} d b \, sin(b) \, \rho(l,b,d)
\end{eqnarray*}
\begin{equation}
\label{eqn:theory2}
\end{equation}
where
\begin{equation}
 F(L_{\gamma},D) = \frac{\Gamma-2}{\Gamma-1} \, \frac{1}{4 \, \pi \, D^2 \, E_{th}} \,L_{\gamma}
\end{equation}

If the luminosity function defined in Eq.(\ref{eqn:luminosity}) has slope $\alpha=-1$
the total flux from resolved and unresolved PWNe divided over the HESS field of view
if 2.5 per cent of radio loud pulsars emit $\gamma$-rays is
\begin{equation}
F_{PWN}(E> 200 GeV, \,  -30<l<30, \, -3<b<3)= 1.3 \times {10}^{-9}
\, photons \, {\mathrm{cm^{-2}}} \, {\mathrm {{sr}^{-1}}} \, {\mathrm {s^{-1}}} \,.
\label{eqn:prima}
\end{equation}
The total flux from the Galactic SNR population is
\begin{equation}
F_{SNR}(E> 200 GeV, \,  -30<l<30, \, -3<b<3) =1.1  \times {10}^{-9} \, photons  \, {\mathrm{cm^{-2}}} \,  {\mathrm {s^{-1}}} \,  {\mathrm {{sr}^{-1}}} \, ,
\end{equation}

For $\alpha=-1$ the integral flux expected for the Milagro field of view from HESS-like sources assuming a pulsar shaped surface density if 2.5 per
cent of radio loud pulsars emit $\gamma$-rays is
\begin{equation}
F_{PWN}(E>200 GeV, \,  40<l<100, \, -5<b<5) =  4.6 \times {10}^{-10} \, photons  \,{\mathrm{cm^{-2}}} \, {\mathrm {s^{-1}}} \, {\mathrm {{sr}^{-1}}} \,.
\label{eqn:terzabis}
\end{equation}
while the contribution from SNRs shaped surface density is
\begin{equation}
F_{SNR}(E>200 GeV, \,  40<l<100 \, -5<b<5) =  4.4 \times {10}^{-10} \, photons  \, {\mathrm{cm^{-2}}} \,  {\mathrm {s^{-1}}} \,  {\mathrm {{sr}^{-1}}} \,.
\label{eqn:quarta}
\end{equation}
The fluxes in Eq.~(\ref{eqn:terzabis}) and (\ref{eqn:quarta}) are lower than the HESS fluxes
from resolved and unresolved sources. In fact, as shown in Fig.\ref{fig3}, HESS observes the inner
region of the Galaxy, whereas Milagro field of view is more spread toward the outer regions
in the Galaxy, where the number of sources is substantially lower, so a lower contribution
from unresolved sources is expected for Milagro's region of the Galactic plane than for HESS.

The integral flux from unresolved pulsars
if 2.5 per cent of radio loud pulsars emit $\gamma$-rays corrected for the Milagro threshold becomes
\begin{eqnarray*}
F_{PWN}(E>3.5 \, TeV, \, 40<l<100) &=&  F_{PWN}(E>200 \, GeV, \, 40<l<100) \\[2mm]
& \times &{(\frac{E_{Milagro}}{E_{HESS}})}^{(-\Gamma+1)} \\[2mm]
&& =  3.6 \times {10}^{-11} \, photons \,{\mathrm{{cm}^{-2}}} {\mathrm {sr}^{-1}} \, {\mathrm {s^{-1}}}
\end{eqnarray*}
\begin{equation}
\label{fluxtotaltris3}
\end{equation}
whereas the contribution from SNRs is
\begin{eqnarray*}
F_{SNR}(E>3.5 \, TeV, \, 40<l<100) &=&  F_{pulsar}(E>200 \, GeV, \, 40<l<100) \\[2mm]
& \times &{(\frac{E_{Milagro}}{E_{HESS}})}^{(-\Gamma+1)} \\[2mm]
&=&   3.4 \times {10}^{-11}  \, photons  \,{\mathrm{{cm}^{-2}}} {\mathrm {sr}^{-1}} \, {\mathrm {s^{-1}}}
\end{eqnarray*}
\begin{equation}
\label{fluxtotaltris2}
\end{equation}
where $E_{Milagro}$ and $E_{hess}$ are the Milagro and HESS energy thresholds,
respectively. Assuming a slope $\alpha=-1.$ for the luminosity functiony and
summing the two contributions from HESS-like SNRs
and PWNe, the emission due to unresolved sources to the Milagro diffuse emission is comparable to
the diffuse emission itself. If the slope of the luminosity function $\alpha=-1.5$ the contribution of unresolved
SNRs and PWNe to the diffuse emission measured by Milagro is about 5 percent and for $\alpha=-2$ this
contribution becomes negligible, which is in disagreement with the lower limit of 10 percent for the
contribution of HESS-like sources to the VHE diffuse emission previously found. Thus the slope of the
luminosity function for HESS-like sources is constrained to be  $-1> \alpha > -1.5$.

\section{Conclusions}

The number-intensity relation and the luminosity function for the HESS source population
were investigated using the assumption that HESS sources are distributed as
PSRs and SNRs detected at radio wavelengths. In order for the chosen sample of sources to be complete
only the HESS sources with fluxes above 6 percent of the Crab flux were taken into account
to derive the number-intensity relation. The contribution of
unresolved HESS-like sources to the diffuse emission measured by Milagro was also estimated.
Using the logN-logS relation for the HESS sample of Galactic $\gamma$-ray emitters
at least 10 per cent of the diffuse emission at TeV energies is estimated to be due to
the contribution of unresolved HESS-like sources. This result is a lower limit for such a contribution
because we have taken into account only sources detected above 6 per cent of the Crab flux and because
HESS sensitivity gets worse for extended sources, meaning that some
extended sources might have been missed by HESS. Using the logN-logS relation
we have also predicted the number of HESS-like sources which VERITAS,
HESS and HAWC should detect during their survey of the sky.
An alternative procedure to evaluate the contribution of unresolved HESS-like sources to
Milagro diffuse emission gives the diffuse flux due to unresolved sources comparable to the
diffuse emission itself. We finally constrained the slope of the luminosity function.
The main uncertainty of this calculation consists in assuming that the distribution of $\gamma$-ray sources follows the distribution of either
pulsars or SNRs observed in the radio. In particular, in order to predict how many
PSRs observed in the radio have a PWN and are possible gamma ray emitters
we used the result that the spin-down energy loss $dE/dt > dE/dt_c = 4 \times {10}^{36} erg/s$ for a young energetic pulsar to form a PWN. In this
respect we have ignored the existence of pulsars, such as Geminga, which are $\gamma$-ray loud,
yet not observed in the radio.

New observational results support the hypothesis that a population of unresolved sources
contribute significantly to the emission at very high energy. Milagro has recently reported the discovery of
TeV gamma ray emission from the Cygnus Region of the Galaxy, which exceeds the predictions of conventional models of gamma -ray production
\citep{Abdo:2006} from the same region in the Galaxy where the Tibet Array has detected an excess of cosmic rays \citep{Amenomori}.
Thanks to its improved sensitivity Milagro has also better imaged the whole Northern sky and discovered four sources and four source
candidates \citep{Abdo:2007}. HESS has seen very high energy emission spatially correlated with giant molecular
clouds located in the Galactic Center \citep{Aharonian:nature}.
The energy spectrum measured by HESS close to the Galactic Center is $E^{-2.3}$, significantly harder than the $E^{-2.7}$ spectrum of the
 diffuse emission and equal to the average spectrum of the HESS source population.
The emission from the Galactic Center might possibly unveil a cosmic ray accelerator.

To draw more definitive conclusions about the very high energy $\gamma$-ray sky, new observations
are of fundamental importance. New hints will be provide by both MAGIC and VERITAS, which already
survey the Cygnus Region. Finally GLAST will investigate the window of energy between 10 MeV to 300 GeV, covering the energy gap left
between EGRET and the ground-based low threshold gamma-ray observatories.


\begin{thebibliography}

\bibitem[Abdo et al. (2007)]{Abdo:2006}
Abdo, A. A. et al, 2007, \apj 658L, L33

\bibitem[Abdo et al. (2007)]{Abdo:2007}
Abdo, A. A. et al, 2007, \apj 664L, 91

\bibitem[Aharonian \& Atoyan (2001)]{Aharonian:2000iz}
Aharonian, F. A. \& Atoyan, A. M., 2001, A\&A, 362, 937

\bibitem[Aharonian et al.(2004)]{Aharonian:2004hegra}
Aharonian, F. A. et al., 2004, \apj 614, 897

\bibitem[Aharonian et al.(2005)]{Aharonian:2005jn}
Aharonian, F. A. et al., 2005, Science 307, 1938

\bibitem[Aharonian et al.(2006a)]{Aharonian:2005kn}
Aharonian, F. A. et al., 2006, \apj 636, 777

\bibitem[Aharonian et al.(2006b)]{Aharonian:nature}
Aharonian, F. A. et al., 2006, Nature 439, 695

\bibitem[Albert et al. (2006)]{Albert:2005kh}
  Albert, J. et al., 2006, \apj 638, L101

\bibitem[Amenomori et al. (2006)]{Amenomori}
  Amenomori, M. et al., 2006, Science 314, 439

\bibitem[Atkins et al.(2000)]{Atkins:2000}
  Atkins, R. W. et al., 2000, Nucl.Instrum Methods Phys. Res., Sect. A, 449, 478

\bibitem[Atkins et al.(2003)]{Atkins:2003}
  Atkins, R. W. et al., 2003, \apj, 595, 803

\bibitem[Atkins et al.(2004)]{Atkins:2004jf}
  Atkins, R. W. et al., 2004, \apj, 608, 680

\bibitem[Atkins et al.(2005)]{Atkins:2005}
  Atkins, R. W. et al., 2005, PRL, 95, 251

\bibitem[Bhattacharya et al.(2003)]{Bhattacharya}
Bhattacharya, D. et al., 2003 A\&A, 404, 163

\bibitem[Bergstrom et al. (2006)]{Bergstrom:2006tk}
  Bergstrom, L., Edsjo, J., Gustafsson, M. and Salati, P., 2006, JCAP 0605, 006

\bibitem[Bertsch et al.(1993)]{Bertsch:1993}
  Bertsch, D. L. et al., 1993, \apj, 416, 587B

\bibitem[Bignami \& Caraveo (1980)]{Bignami}
  Bignami, G. F. \& Caraveo, P. A., 1980, \apj, 241, 1161

\bibitem[Case \& Bhattacharya (1998)]{Case:1998qg}
Case, G. L. \& Bhattacharya D., 1998, \apj, 504, 761

\bibitem[Combi et al. (2001)]{Combi}
Combi, J.A. et al., 2001, A\&A, 366, 1047

\bibitem[de Boer et al.(2005)]{deBoer:2005tm}
de Boer, W., Sander, C., Zhukov, V., Gladyshev, A. V. and Kazakov, D. I., 2005, A\&A, 444, 51

\bibitem[Esposito et al. (1996)]{Esposito}
Esposito, J.A. et al., 1996, \apj 461, 820

\bibitem[Funk (2007)]{Funk}
  Funk, S., 2007, Astrophysics and Space Science, 309, Issue 1-4, 11

\bibitem[Green (2004)]{Green:2004gr}
  Green, D. A., 2004, \,Bull.\ Astron.\ Soc.\ India 32, 335

\bibitem[Grimm, Gilfanov \& Sunyaev (2002)]{Grimm}
Grimm, H. J, Gilfanov, M. \& Sunyaev, R., 2002, A\&A, 391, 923

\bibitem[Gonthier et al.(2002)]{Gonthier}
 Gonthier et al., 2002, \apj 565, 482

\bibitem[Gotthelf (2003)]{Gotthelf:2003}
  Gotthelf, E. V., 2003, \apj, 591, 361

\bibitem[Gotthelf (2004)]{Gotthelf:2004}
  Gotthelf, E. V., 2004 IAU Symposium, 218, 225G

\bibitem[Hunter et al.(1997))]{Hunter:1997we}
  Hunter, S. D. et al., 1997, \apj, 481, 205

\bibitem[Kamae et al.(2006))]{Kamae:2006}
 Kamae, T. et al., 2006, \apj, 647, 692

\bibitem[Lorimer et al. (2004)]{Lorimer:2004}
Lorimer, D. R. et al., 2004, IAU Symposium, 218, 105

\bibitem[Lorimer et al. (2006)]{Lorimer:2006}
Lorimer, D. R. et al., 2006, MNRAS, 372, 777

\bibitem[Manchester et al.(2005))]{Manchester}
 Manchester, R. N. et al., 2005, Astron. Journ., 129, 1993

\bibitem[Manchester (2005))]{Manchester:2004}
 Manchester, R. N., 2005, Ap\&SS., 297, 101

\bibitem[Muslimov \& Harding (2003)]{Muslimov}
Muslimov, A.G. \& Harding, A.K., 2003, \apj 588, 430

\bibitem[\"Ozel \& Thompson (1996)]{Oezel}
 \"Ozel, M. E \% Thompson, D. J., 1996, A\&A, 463, 105

\bibitem[Prodanovic et al.(2006)]{Prodanovic}
  Prodanovic, T. et al., 2007, Astroparticle Physics 27, 10

\bibitem[Romero et al. (1999)]{Romero}
Romero, G.E., Benaglia, P. \& Torres, D.F., 1999, A\&A, 348, 868

\bibitem[Smith et al.(2005) ]{Smith}
	Smith, A. J. et al., 2005  Symposium on High-Energy Gamma-Ray Astronomy,
 AIP Conference Proceedings, 2005, 745, 657

\bibitem[Stecker, Hunter \& Kniffen (2007) ]{Stecker}
Stecker, F. W., Hunter, S. D. \&  Kniffen, D. A, 2007, astro-ph 0705.4311v3, Astroparticle Physics (in press)

\bibitem[Strong et al.(1999)]{Strong:1999hr}
Strong, A. W., Moskalenko, I. V. \& Reimer O., 1999 \apj, 537, 763

\bibitem[Strong et al.(2004)]{Strong:2004}
Strong, A. W., Moskalenko, I. V. \& Reimer O., 2004 \apj, 613, 962

\bibitem[Strong et al.(2005)]{Strong:2005}
Strong, A. W. et al., 2005 A\&A, 444, 495

\bibitem[Strong (2006)]{Strong:2006}
Strong, A. W., 2006, Proceedings of Conference 'The multi-messenger approach to high-energy gamma-ray sources', Barcelona, 2006

\bibitem[Sturmer \& Dermer (1995)]{Sturmer}
Sturmer, S.J. \& Dermer, C.D., 1995, A\&A, 293,L17

\bibitem[Swordy (2003)]{Swordy}
Swordy, S., 2003, 28th International Cosmic Ray Conference, Universal Academy Press Inc.

\bibitem[Xu et al. (2005)]{Xu}
Xu,J.,Zhang,X.\& Han,J., 2005, Chin. J. Astron. Astrophys, 5, 165

\bibitem[Yusifov \& K\"{u}\c{c}\"{u}k (2004)]{Yusifov:2004fr}
 Yusifov, I. \&  K\"{u}\c{c}\"{u}k, I., 2004, A\&A, 422, 545

\bibitem[Zhang et al.(2004)]{Zhang}
Zhang, L. et al, 2004, \apj 604, 317

\end{thebibliography}
\end{document}